\documentclass[conference]{IEEEtran}

\usepackage[utf8]{inputenc}
\usepackage{acronym}
\usepackage{multirow}
\usepackage{soul}
\usepackage{hyperref}
\usepackage{algorithm2e}
\usepackage{listings}
\usepackage{array}
\usepackage{tabularx}
\usepackage{svg}
\usepackage{amsmath,amssymb,amsfonts}
\usepackage{pifont}
\usepackage[normalem]{ulem}
\usepackage{flushend}

\usepackage[
    backend=biber,
    sorting=none,
    citestyle=numeric,
    bibstyle=numeric,
    maxbibnames=99,
]{biblatex}
\title{Auditable data structures: theory and applications}

\author{
    \IEEEauthorblockN{Andrea Canciani}
    \IEEEauthorblockA{
        \textit{Geckosoft}\\
        a.canciani@geckosoft.it
    }
    \and
    \IEEEauthorblockN{Claudio Felicioli}
    \IEEEauthorblockA{
        \textit{Traent}\\
        claudio.felicioli@traent.com
    }
    \and
    \IEEEauthorblockN{Fabio Severino}
    \IEEEauthorblockA{
        \textit{Traent}\\
        fabio.severino@traent.com
    }
    \and
    \IEEEauthorblockN{Domenico Tortola}
    \IEEEauthorblockA{
        \textit{University of Pisa}\\
        domenico.tortola@phd.unipi.it
    }
}

\begin{document}

\maketitle

\thispagestyle{plain}
\pagestyle{plain}

\begin{abstract}
    Every digital process needs to consume some data in order to work properly. It is very common for applications to use some external data in their processes, getting them by sources such as external APIs. Therefore, trusting the received data becomes crucial in such scenarios, considering that if the data are not self-produced by the consumer, the trust in the external data source, or in the data that the source produces, can not always be taken for granted. The most used approach to generate trust in the external source is based on authenticated data structures, that are able to authenticate the source when queried through the generation of proofs. Such proofs are useful to assess authenticity or integrity, however, an external user could also be interested in verifying the data history and its consistency. This problem seems to be unaddressed by current literature, which proposes some approaches aimed at executing audits by internal actors with prior knowledge about the data structures. In this paper, we address the scenario of an external auditor with no data knowledge that wants to verify the data history consistency. We analyze the terminology and the current state of the art of the auditable data structures, then we will propose a general framework to support external audits from both internal and external users.
\end{abstract}


\section{Introduction}\label{sec:introduction}
\noindent Nowadays, data are the engine for every digital process. Every category of digital activity, at some point, needs to process or evaluate some kind of data to be completed. In real-world scenarios, applications often exchange data with external sources, such as cloud service providers. This highlights the need by the user of such services to trust the received data and the operations executed on that data.

There are well-established lines of research focused on data integrity and authentication when the data publisher or the outsourced data manager are untrusted, proving that the data has not been tampered with. On the other hand, the current literature that addresses the auditability of the sequence of operations executed on the data is at an early stage, even if very promising.

The lack of standard definitions for key concepts such as data consistency and data history is the main motivation behind this paper. Our contributions are the proposal of some terminology and of a new framework, that can be useful when describing and comparing solutions in the line of research of data structure history auditing. We are also contributing with a small survey about a selection of existing solutions (including some commercial products not yet described in scientific literature) that make use of auditable data structures to improve the overall trust not only in the data but also in the process, considered as the sequence of operations that leads to that data.

The paper is structured as follows: in the rest of this Section, we will introduce some terminology and key concepts, then we will use them to describe two authenticated data structure models commonly adopted in the scientific literature when addressing data replication and data outsourcing scenarios; in Section \ref{sec:framework} we will describe the scenario of external audit of history consistency, and we will propose a framework to address it; in Section \ref{sec:applications} we will discuss the current state of the art for data structure history consistency auditing, focusing on a few notable applications; Section \ref{sec:conclusion} concludes the paper.

\subsection{Terminology}\label{sec:terminology}

\noindent For the sake of clarity, here we introduce some terminology and key concepts that will be largely used in the next sections. This terminology will be useful to fully understand the objectives and the underlying ideas behind the externally auditable data structure framework that we are proposing in Section \ref{sec:framework}. Some terms also apply to the context of authenticated data structure models.

\begin{itemize}
    \item Data structure \textbf{operation}: a function defined on the data structure. It can be executed, by taking arguments and returning a result. If the execution can have the side effect to modify the state of the data structure, then it is an \textbf{edit operation}, otherwise, it is a \textbf{query operation}.
    \item Data structure \textbf{digest}: an identifier of the current state of the data structure. It is usually compact and computed by cryptographic hash composition. In order to be safely used in the context of authenticated or auditable data structures, it must be at least second-preimage resistant (i.e., given a digest derived from a certain data structure state, it should be computationally hard to find a different state identified by the same digest value).
    \item Data structure \textbf{consistency}: the property of two time-ordered data structure instances where the newer one can be produced starting from the older one by executing a sequence of valid edit operations.
    \item Data structure \textbf{history}: a data structure that identifies a chronologically ordered sequence of versions of the same data structure. For instance, a simple data structure history can be a list of digests.
    \item \textbf{History consistency}: the property of data structure history where consistency holds for every pair of data structure versions.
    \item \textbf{Proof}: a statement, paired with some supporting data. The supporting data must be sufficient to independently verify that the statement is true with a high probability, without relying on trust.
    \item \textbf{Integrity proof}: a proof stating that a given operation executed on a data structure with a given digest will produce a given result. Inclusion proofs (for example, the ones implemented as the hash path from the root to the leaf of a Merkle tree) are a special case of integrity proofs.
    \item \textbf{Consistency proof}: a proof stating that two time-ordered data structures with given digests are consistent.
    \item \textbf{Audit}: an inspection of a data structure aimed to check some properties of the data structure content in present or past times, and/or some properties of the sequence of edit operations executed on it.
    \item \textbf{External auditor}: an auditor without prior knowledge of the data structure.
    \item \textbf{Externally auditable data structure}: a data structure whose history consistency can be verified by an external auditor without compromising data secrecy (i.e. the history must be devoid of data, for example by using one-way data indirections).
\end{itemize}

\subsection{Related works}\label{sec:relatedWorks}

\noindent The historical roots of the externally auditable data structure framework that we are introducing can be found in the line of research of the authenticated data structures; they have first been proposed as a solution to safely distribute certificate revocation lists \cite{laurieCertificatesRevocation}. The use of authenticated data structures, alongside with a verifiable set of operations which includes insertion and deletion, makes the entire revocation process more transparent, improving the auditability of the revocation list.

In \cite{devanbu2000authenticated} the first formal model for authenticated data structures, nowadays known as the \textit{three-party model}, was proposed, addressing the data replication scenario.

\begin{figure}[htbp]
    \centering
    \includegraphics[width=\columnwidth]{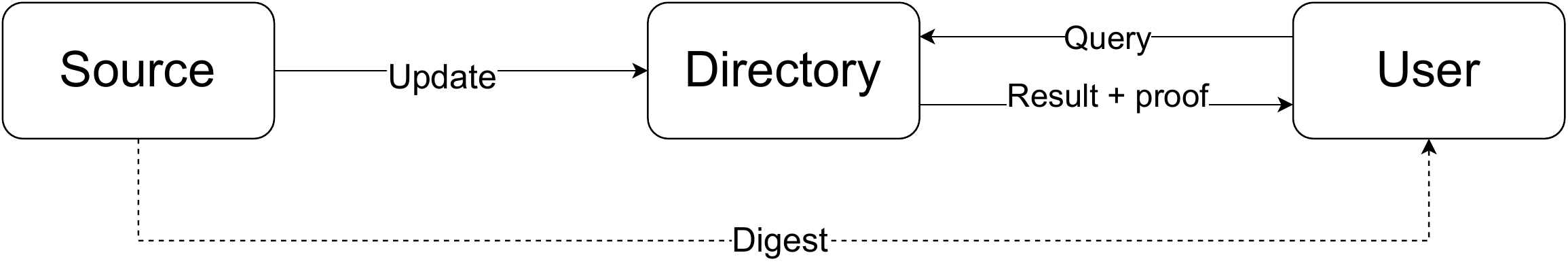}
    \caption{Scheme of the three-party model for data replication}
    \label{fig:3party}
\end{figure}

In the data replication scenario, a data structure is managed by a trusted \textit{source} and replicated by an untrusted \textit{directory} to delegate distribution and also improve scalability. When querying the \textit{directory}, the \textit{user} needs a way to check that responses have not been tampered with. In the three-party model, schematized in Figure \ref{fig:3party}, the \textit{user} knows a digest of the data structure authenticated by the \textit{source}, and the \textit{directory} produces an integrity proof when responding to queries.

To be utilized in the three-party model, a data structure needs to meet certain key requirements:

\begin{itemize}
    \item It must possess a method for computing a second-preimage resistant digest, which identifies the current state of the data structure.
    \item For each query operation intended for use by the \textit{user}, an algorithm must be defined that generates integrity proofs. Additionally, a corresponding proof verification algorithm should be defined.
\end{itemize}

The three-party model has been adopted by several works as a common framework, useful to describe and compare a large repertoire of authenticated data structures \cite{tamassia2003Authenticated, anagnostopoulos2001persistent, goodrich2001implementation, reyzin2017blockchain}

In \cite{papamanthouT20072parties} the three-party model has been extended to the more general \textit{two-party model}, addressing the data outsourcing scenario, an increasingly common scenario in the context of cloud services.

\begin{figure}[htbp]
    \centering
    \includegraphics[width=\columnwidth]{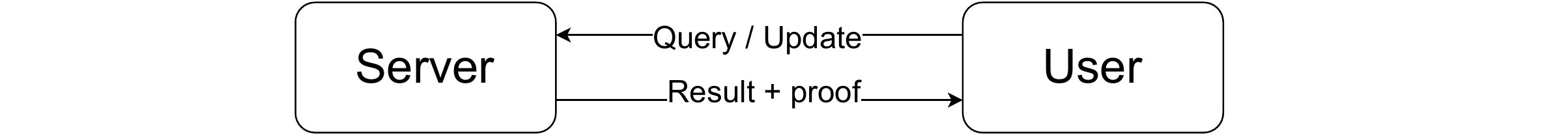}
    \caption{Scheme of the two-party model for data outsourcing}
    \label{fig:2party}
\end{figure}

In the data outsourcing scenario, the \textit{user} delegates the management of a data structure to an untrusted \textit{server} while retaining the ownership. The \textit{user} is not required to maintain a replica of the data structure and needs a way to check that it has not been tampered with. In the two-party model, schematized in Figure \ref{fig:2party}, the \textit{user} is assumed to know the most recent digest of the data structure; the \textit{server} produces an integrity proof when responding to queries, and a consistency proof when responding to edit operation requests, stating the updated value for the data structure digest.

The key requirements for a data structure to be utilized in the two-party model are essentially the same as those for the three-party model, with an additional one: for every edit operation, an algorithm must be defined that generates consistency proofs, and a corresponding proof verification algorithm should also be defined. This indicates that the requirements for the three-party data replication model are encompassed within those of the two-party data outsourcing model. Hence, the two-party model can be seen as a generalization of the three-party model.

\section{Externally auditable data structure framework}\label{sec:framework}

\begin{figure*}[htbp]
    \centering
    \includegraphics[width=0.8\paperwidth]{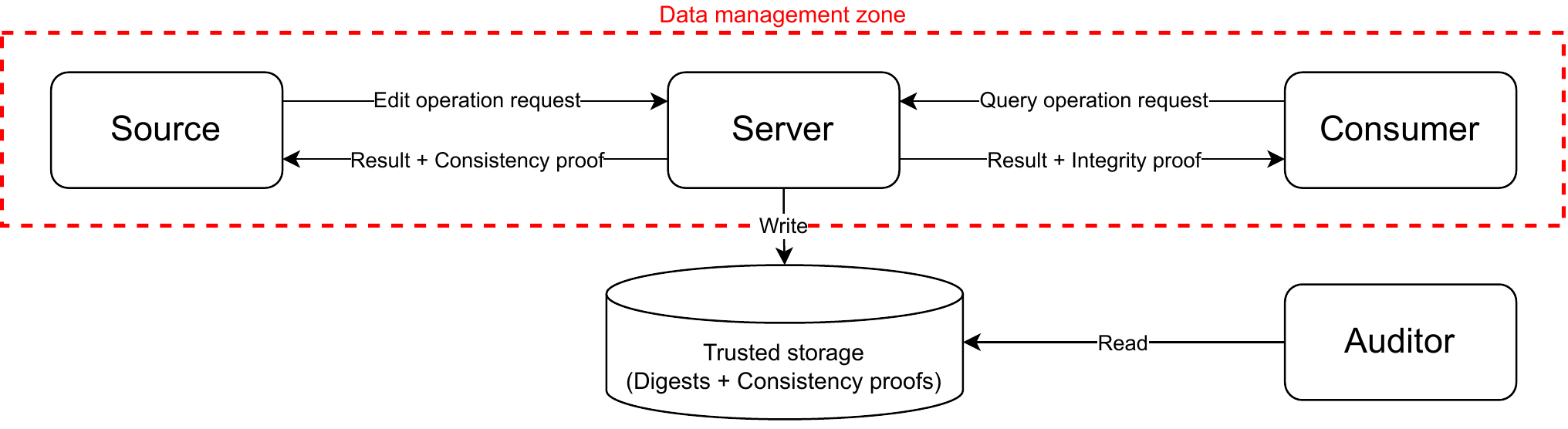}
    \caption{Scheme of the externally auditable data structure framework}
    \label{fig:generalModel}
\end{figure*}

\noindent Now we describe the scenario of external audit of history consistency, a data secrecy preserving scenario that extends the two-party authenticated data structure model, for which we propose the externally auditable data structure framework, schematized in Figure \ref{fig:generalModel}.

In this scenario, the \textit{source} delegates the management of a data structure to an untrusted \textit{server} while retaining the ownership. The \textit{consumer} query the untrusted \textit{server}. The external \textit{auditor} verifies the history consistency of the data structure managed by the \textit{server}.

In the externally auditable data structure framework, the \textit{server} produces an integrity proof when responding to queries, and a consistency proof when responding to edit operation requests, stating the updated value for the data structure digest. The \textit{server} also writes to the trusted \textit{storage} an authenticated data structure history, including a proof stating its consistency, extending it incrementally whenever an edit operation is executed. The external \textit{auditor} verifies history consistency thanks to the data from the trusted \textit{storage}, including the digest list and the consistency proofs between each pair of consecutive digests. Therefore, there is no need to interact with the \textit{server}.

The key requirements for a data structure to be utilized in this framework are essentially the same as those for the two-party model, with the additional one that the consistency proofs must preserve data privacy. Hence, the externally auditable data structure framework can be seen as a generalization of the two-party data outsourcing model.

\section{Notable applications}\label{sec:applications}

\noindent In this section, an overview of the applications of auditable data structures will be discussed. Such applications include digital certificate management (release and revocation), database query verification, log file management, and private distributed ledger auditing.

The mentioned applications often implement the three-party data replication model or the two-party data outsourcing model, analyzed in Section \ref{sec:relatedWorks}. The discussion proposed in the next paragraphs will be focused on how these implementations can handle internal and external auditing and on the data exposed during the audit process. The three approaches most commonly adopted to enable auditing are: edit operations replay, data-exposing proofs, and Merkle proofs.

In the edit operations replay approach, an auditor is able to verify any property of the edit operations executed on a data structure in the most straightforward way possible: the full list of edit operations is made available and re-executed on an older version of the data structure already trusted by the auditor, then the identity between the resulting version and the one under audit is checked. Proof-based approaches are more elegant and efficient, but usually it is hard not to expose any data in the process. Merkle proofs appear to be the most suitable proving scheme when addressing use cases with the requirement of data secrecy preservation since they consist only of hashes. Merkle proofs make use of Merkle trees \cite{merkleTreeSurvey} as an underlying data structure that enables the construction of both integrity (inclusion) and consistency proofs. The former is generated in several cases, such as when a log is queried to verify the presence of a certain entry or in authenticated databases in order to prove the presence of a certain record; the latter can be generated, for example, to prove that a newer version of a log is consistent to a trusted older version, meaning that all the entries from the older one are preserved, in the same order, and are a prefix for the entry list of the newer log.

\subsection{Digital certificate management systems}\label{sec:digitalCertificates}

\noindent According to \cite{naori2000certificates} digital certificate management system can be modeled as a three-party data replication model (see Section \ref{sec:relatedWorks}) where the source is represented by the certification authority (CA), directories are one or more untrusted parties that serve as certificate databases to users in order to avoid bottlenecks with direct accesses to the source, and the users get their certificates and issue queries to the publishers to get certificate information.

Laurie and Kasper provided an innovative way to use verified data structures to handle certificate transparency \cite{laurieCertificatesTransparency} and revocation \cite{laurieCertificatesRevocation}, with the final goal to make the most difficult as possible to a CA to release a certificate for a domain without the domain owner to know about that. Certificates are managed as a log file, where every certificate is appended when released. When required, certificates have to be accompanied with an integrity proof in one or more of these log files. Logs need to be consistent, meaning that every log must contain every certificated signed by the log itself. Logs are defined as a Merkle tree structure containing signed certificates as leaf nodes, and periodically (according to a time parameter) logs produces a digest made by the Merkle root, the number of entries, a timestamp and the log digital signature. CAs are also required to be able to provide integrity proof for each entry every time an auditor issues a query for it. Such integrity proof is a classical Merkle inclusion proof made by the path from the root to the corresponding leaf node.

Regarding the certificate revocation, the authors introduced the deletion operation to the mechanism used with the certification transparency. Logs still have to be able to prove the consistency of the history, meaning that a certificate could be or not be present in a particular version of the log, keeping a valid history for itself between each version. This can be done with a Merkle tree where every leaf node contains pairs from a sorted list of revoked certificates, or by using sparse Merkle trees \cite{dahlberg2016sparseMerkleTree} to store the status on every possible certificate.

\subsection{Log management systems}\label{sec:logs}

\noindent At a high level, a log file can be modeled as a Merkle tree where every new event captured by the log is a leaf node. To guarantee that the log works properly, the appending of a new element (which equals to a new event captured by the log) is the only valid operation. With such configuration, a log can be used as an externally auditable data structure, and the consistency between any two version of the same log file can always been proved by computing Merkle consistency proofs. Auditability is a key property for a log management system in order to improve transparency and to guarantee the tamper-evidence of the log itself \cite{bao2020log}.

In \cite{ejidenberg2015verifiable}, three verifiable data structures that can be used to successfully implement a verifiable log management system are proposed:

\begin{itemize}
    \item \textbf{Verifiable log}: a verifiable log is an append-only tree data structure designed to be used, as the name suggests, for log management. This persistent data structure starts empty and generates a new version of itself every time a new row is appended. Periodically, an authenticated digest is published, which includes the tree root hash value, the tree size, and a digital signature. The tree can be queried to check entry inclusion, return all the entries in the tree, and verify the append-only property.
    \item \textbf{Verifiable map}: a verifiable map is a tree data structure where leaf nodes are key-value pairs. Besides the differences between stored data format, verifiable maps work similarly to the verified log and can be implemented as a sparse Merkle tree \cite{dahlberg2016sparseMerkleTree}. Periodically, an authenticated digest is generated from an instance of the map, including the root hash and a digital signature. A verifiable map can be queried to verify the inclusion of a certain key, to retrieve a value by key, and to enumerate all the key-value pairs.
    \item \textbf{Log-backed map}: this data structure is a combination of a verifiable log and a verifiable map: the map is populated with key-value pairs, while the log describes an ordered list of edit operations, which can be replayed in order to verify the history consistency of the map. For this data structure, the digest production has to keep track of the log version related to the specific map version. This hybrid design address the verification of data structure consistency between two different versions of the map by adopting the edit operation replay approach (using the operations included in the log), while the log supports data structure consistency verification by adopting the Merkle proof approach.
\end{itemize}

Trillian\footnote{\url{https://transparency.dev/\#trillian} [Accessed on 22 May 2023]} is a commercial log management provider that uses the aforementioned verified data structures \cite{trillianWebsite} to implement a log system, with the goal of guaranteeing log immutability and to provide an auditable event history. 

The Trillian design allows the creation of integrity proofs for query results, and consistency proofs as explained in Section \ref{sec:framework}, making it possible to externally verify the data structure consistency between consecutive versions of the log.

\subsection{Auditable database providers}\label{sec:db}

\noindent The main role of auditability for databases is to create a tamper-evidence database, generating inclusion proofs for each record included in any query response, and also consistency proofs to verify the transaction history.

LedgerDB \cite{yang2020ledgerdb} is a centralized database service that offers tamper-evidence, non-repudiation and strong auditability. In LedgerDB, auditability is guaranteed by using a data structure called \textit{journal}, related to the database, and by implementing a transaction execution flow that ensures persistency and tamper-resistance of the processed transactions. When an internal client requests an audit, the journal is populated with client-related data fields (such as client ID, nonce and timestamp) and operator related fields (URI, type, transaction). A hash for the request is calculated according to such data and signed by the client. Then the database server adds some fields such as the request hash, the data stream location and a timestamp. Finally, the server calculates the entire data structure hash, sign it and returns to the client as a \textit{journal receipt}. This protocol ensures that an internal client can always verify the operation validity too, using the journal sequence as an operation log. On the other hand, the transaction execution flow is divided into four phases: \textit{verification} (of the transaction validity), \textit{execution}, \textit{commit} (journal data structure generation and processing) and \textit{indexing} (index generation for the committed transaction, in order to optimize data retrieval and verification). With this setup, LedgerDB supports external audits for transaction inclusion, notary time anchors and verified data removal.

By requiring access to the journal and its underlying data, accordingly to the definitions provided in Section \ref{sec:terminology}, an auditor must be internal to the system. Therefore, LedgerDB does not support external auditability of history consistency of the database.


QLDB (\textit{Quantum Ledger Database}) \cite{qldbWebsite} is a document based ledger database developed by Amazon that provides immutable, tamper-proof transaction log useful to track database changes and to generate a unique and complete change history. The database is backed with an append only log, meaning that every change is verifiable and auditable. Users can ask for the document digest, calculated as a SHA-256 hash of the document, and use it to verify the document correctness. It is also possible to verify the document history to check the data history. This setup allows easy internal audits, and also external audits with the drawback to share data with the auditor.

\subsection{Auditable private distributed ledger systems}\label{sec:traent}

\noindent Blockchain technology, born with Bitcoin \cite{Bitcoin}, builds a P2P network whose peers do not know and do not trust each other but work together to maintain a shared state (the ledger of all the transactions between the users). A consensus algorithm defines the rules to upgrade the state, e.g. by deciding which peer will compute and communicate the next chunk of data \cite{ConsensusSurvey}. The generalization of this technology is also known as Distributed Ledger Technology (DLT), since not all implementations use a chain of blocks (e.g., IOTA utilizes a Direct Acyclic Graph called Tangle \cite{Iota}).

A DLT is public and permissionless when any peer can send and receive the transactions, and join the consensus without restrictions. Private and permissioned DLTs have been developed to address different use cases, shifting the attention to smaller scale system where the peers do know but do not trust each other \cite{NeedBC}.

Public and permissionless DLTs guarantee high security, decentralization and provide transparent systems; however, they suffer of scalability problems (causing a low transaction throughput), require fees to execute transactions (which can result very expensive in some networks), and do not comply with privacy-preserving regulations. Instead, private and permissioned DLTs have higher throughput, lower infrastructural costs, and are able to limit data access. On the other hand, they are less decentralized and the security properties apply only to the internal participants \cite{NeedBC}. Typically, for industrial use cases the trend is to adopt private and permissioned DLTs \cite{Polge}, with Hyperledger Fabric \cite{HFabric} being one of the most popular frameworks used to build them.

In the case of public DLTs, where there is no distinction between internal and external, auditability derives naturally from the immutability and transparency properties \cite{BlockchainACMaesa}. Conversely, in the case of private DLTs there are known issues preventing external auditability: when a piece of data from a private DLT is made public for the first time, is not possible for somebody outside of the network to verify its consistency with the contents of the private ledger due to access limitations. More critically, even when full access to the private network is granted to an auditor, it is still possible for the participants of the private network to agree to re-write history right before the auditor would join the network, effectively forking the ledger (a history re-writing fork has been executed publicly on Ethereum in response to the DAO attack \cite{Dao}, a fork can be done more easily, and secretly, on a smaller and controlled network \cite{consortium}). This problem is present also when a new node joins the private network.

Traent\footnote{\url{https://traent.com/} [Accessed on 22 May 2023]} hybrid blockchain \cite{hybridDlt} is an enterprise blockchain solution where private ledgers produced on-demand inside a private network are notarized using a public blockchain as the trusted storage, in order to obtain external auditability of the history consistency while preserving data secrecy.

According to the externally auditable data structure framework modeled in Section \ref{sec:framework}, the Traent hybrid blockchain network covers all the data management zone, with nodes in the private network acting as both source and consumer (they are coordinated to produce and query the data in collaboration with other peers), while the notarization system has the role of server. An auditor component, known as a \textit{monitor}, is available and can be operated outside of the private network. The notarization system periodically updates the notarized data structure history of a specific private ledger by publishing on a public blockchain (\textit{Algorand}\footnote{\url{https://developer.algorand.org/} [Accessed on 22 May 2023]} by default, but the implementation is agnostic) the digest of the current version of the ledger paired with the consistency proof between it and the previously published digest.

When a ledger history is notarized, it is possible to verify several properties, such as the integrity and authenticity of the data structure or a portion of it, its history consistency (fork detection), and the uniqueness of the history. By implementing all the elements of the externally auditable data structure framework, Traent make it possible for an auditor to verify that the only edit operations executed on a ledger are the allowed ones (i.e. that the ledger data structure has a unique consistent history) without requiring auditors to be inside the private network or to have access to any data block content, hence the ledger data remain undisclosed.

In order to scale the notarization to an increasing number of managed ledgers and minimize the amount of data to be written on a public blockchain, Traent employs another data structure, called ledgers tree, to identify with a single sequence of digests the data structure history of every ledger included in a persistent collection. By writing to the public blockchain that sequence of digests, it is possible to produce a proof stating the history consistency for any specific ledger included in the collection; this kind of proof can be verified externally using just the ledgers tree sequence of digests and preserving the privacy of the data contained in the ledger.

\section{Conclusion}\label{sec:conclusion}

\noindent In this paper, we proposed some formal definition for terminology and key concepts pertinent to the field of authenticated and auditable data structures. These definitions aim to facilitate the functional comparison of solutions within these lines of research, focusing on the different approaches and features rather than quantitative or performance-related aspects. We discussed the data replication and outsourcing scenarios, describing the existing models (the three-party model addressing data replication and the two-party model addressing data outsourcing) before proposing a new, more general, framework for external auditability of data structures, that extends the authenticated data structure models to address the scenario of data secrecy preserving external audit of history consistency. We surveyed the current uses of auditable data structures, focusing on a selection of applications such as digital certificates management, log management, auditable database providers, and auditable private DLT systems.

As a future work, we plan to describe in details the ledgers tree data structure discussed in \ref{sec:traent} and the properties of the correlated history consistency proofs. Also, we are planning to discuss the potential benefits of the adoption of the externally auditable data structure framework in very specific real world use cases, as the implementation of a privacy preserving digital product passport.

\section{Acknowledgements}
Our heartfelt thanks go to Professor Laura Ricci and the dedicated researchers of the Pisa Distributed Ledger Laboratory\footnote{\url{https://sites.google.com/unipi.it/pisadltlaboratory/} [Accessed on 22 May 2023]} for the invaluable insights and constructive feedback. We are also deeply grateful to Professor Paolo Ferragina, whose meticulous review of this manuscript has considerably enhanced the quality of our paper. We truly appreciate the time, effort, and expertise they have invested in helping us refine our research.

\section{Disclaimer}
This research is sponsored by Traent. An embodiment of the externally auditable data structure framework applied to the distributed ledger technology, as well as the notarization systems described in \ref{sec:traent}, are patents pending by Traent.

\printbibliography

\end{document}